\documentclass{moriond}

\usepackage[T1]{fontenc}
\usepackage{amsmath}
\usepackage{graphicx}
\usepackage{epstopdf}
\usepackage{bm}
\usepackage{epsfig}
\usepackage{graphics}
\usepackage{xspace}
\usepackage{hyperref}
\usepackage{slashed}
\usepackage{xcolor}
\usepackage{longtable}
\usepackage[small]{subfigure}
\usepackage[normalem]{ulem}
\usepackage[symbol]{footmisc}

\def\be{\begin{equation}}
\def\ee{\end{equation}}
\def\bea{\begin{eqnarray}}
\def\eea{\end{eqnarray}}

\newcommand{\Photo}{}

\begin{document}

\begin{flushright}
MIT/CTP-5723 \\
CERN-TH-2024-063\\
UWThPh 2024-12\\
\end{flushright}
\vspace{0.5cm}

\title{Determining $\alpha_s(m_Z)$ from Thrust with Power Corrections \footnote{Contribution to the proceedings of the 58$^{\rm th}$ Recontres de Moriond 2024, QCD and High Energy Interactions.}}

\author{Miguel A. Benitez-Rathgeb\,\footnote[2]{Speaker}$^{,a}$, Andr\'e H.\,Hoang$^b$, Vicent Mateu$^a$, Iain W.\,Stewart$^c$, Gherardo Vita$^d$}

\address{\flushleft \quad $^a$Dpto.\ de F\'isica Fundamental e IUFFyM, Universidad de Salamanca, E-37008 Salamanca, Spain\\
\quad $^b$University of Vienna, Boltzmanngasse 9, A-1090 Wien, Austria\\
\quad $^c$Center for Theoretical Physics,\,Massachusetts Institute of Technology,\,Cambridge,\,MA\,02139,\,USA\\
\quad $^d$CERN, Theoretical Physics Department, CH-1211 Geneva 23, Switzerland}

\maketitle\abstracts{We update and extend a previous N$^3$LL$^\prime$\,+\,${\cal O}(\alpha_s^3)$ strong coupling determination from thrust data. In particular, we carry out a fit with data fully restricted to the dijet region seeking to minimize the potential impact of power corrections that go beyond dijet configurations. In addition, we parametrize deviations from the dijet power correction in order to add an additional source of uncertainty in the result for $\alpha_s(m_Z)$. We also show that the inclusion of resummation is important to achieve stability with respect to varying the fit region.}

\vspace*{-1cm}

\section{Introduction}\label{sec:Intro}

There exists a discrepancy between strong coupling determinations from $e^+e^-$ event-shape data based on QCD methods to parametrize and fit power corrections~\cite{Abbate:2010xh} and the world average~\cite{Workman:2022ynf}. This discrepancy triggered a computation of power corrections in the 3-jet region based on renormalon analysis~\cite{Caola:2021kzt,Caola:2022vea}, which models how the leading power correction $\Omega_1$ varies across the spectrum~\cite{NZcomment}. Based on this study, a general assessment of strong coupling fits from event-shape data was carried out by the authors of Ref.~\cite{Nason:2023asn}, in which it was concluded that the uncertainties on $\alpha_s$ due to the treatment of non-perturbative power corrections should increase.
In these proceedings, we conservatively assess different sources of uncertainty that impact fits for $\alpha_s(m_Z)$ from thrust, much in the spirit of Ref.~\cite{Nason:2023asn}, and provide an updated fit for $\alpha_s(m_Z)$ concentrating on the dijet region. An article describing our findings in more detail, which also includes full results for a new fit for $\alpha_s(m_Z)$, is in preparation~\cite{Benitez:2024}.

\section{Dijet Factorization Formula}\label{sec:factThm}

Our theoretical description of the event-shape variable thrust $T=1-\tau$ is based on a factorization formula from Soft-Collinear Effective Theory~\cite{Abbate:2010xh,Fleming:2007qr,Schwartz:2007ib}
\begin{equation}
\label{eq:factFormula}
\frac{{\rm d} \sigma}{{\rm d} \tau} = \int {\rm d} k \biggl( \frac{{\rm d} \hat{\sigma}_{\rm s}}{{\rm d} \tau} + \frac{{\rm d} \hat{\sigma}_{\rm ns}}{{\rm d}\tau} \biggr)\! \biggl(\tau - \frac{k}{Q}\biggr) F_\tau(k)\,,
\end{equation}
that consists of three main parts: the singular contribution for massless quarks, $\frac{{\rm d} \hat{\sigma}_{\rm s}}{{\rm d} \tau}$, which dominates in the peak and tail, for which resummation is implemented, the non-singular contribution, $\frac{{\rm d} \hat{\sigma}_{\rm ns}}{{\rm d}\tau}$, which encodes kinematically power-suppressed contributions, and the non-perturbative soft function, dubbed shape function $F_\tau(k)$, that parametrizes the dominant source of hadronization effects. In the tail region of the thrust distribution, which is dominated by dijet and back-to-back collimated 3-jet final states, one is able to perform an operator expansion of the shape function in powers of $\Lambda_{\rm QCD}/(Q\tau)$, where the leading non-perturbative matrix element is $\Omega_1$. This approach is what is referred to as the dijet treatment of the power corrections.

\subsection{Improvements in Theory Description}\label{subsec:improvements}

\begin{figure}
\begin{minipage}{0.33\linewidth}
\centerline{\includegraphics[width=0.87\linewidth]{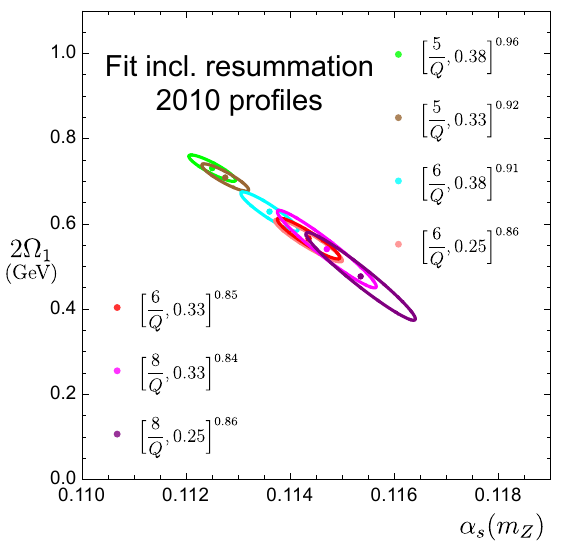}}
\end{minipage}
\hfill
\begin{minipage}{0.32\linewidth}
\centerline{\includegraphics[width=0.87\linewidth]{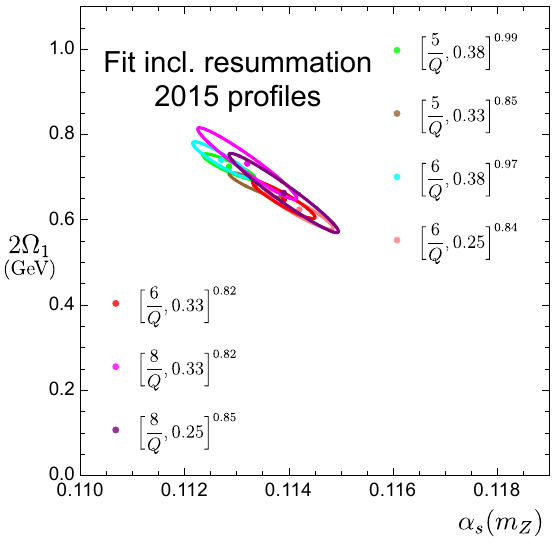}}
\end{minipage}
\hfill
\begin{minipage}{0.32\linewidth}
\centerline{\includegraphics[width=0.87\linewidth]{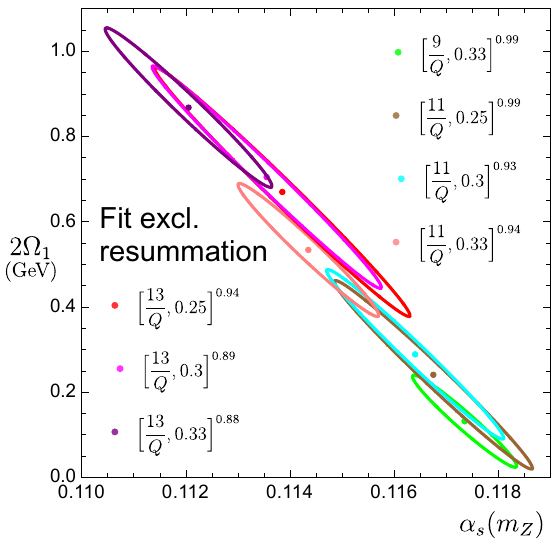}}
\end{minipage}
\caption[]{Results of two-parameter fits for different fit ranges. The ellipses contain only the experimental uncertainties. The intervals and the corresponding superscripted numbers correspond to: $[\tau_{\min},\tau_{\max}]^{\hat \chi^2}$ with $\hat \chi^2=\chi^2/{\rm dof}$. The left panel corresponds to a setup using the 2010 profile functions, the center panel uses the updated 2015 profiles and the right panel does not include resummation at all.}
\label{fig:stability}
\end{figure}

The singular contribution to the total cross section given in Eq.~\ref{eq:factFormula} is governed by three renormalization scales: the hard scale $\mu_H$, the jet scale $\mu_J$ and the soft scale $\mu_S$. The dependence of these scales on $\tau$ in different regions, and transitions between them, is described by so-called profile functions, which have been updated compared to the 2010 analysis~\cite{Abbate:2010xh} by using the 2015 results from Ref.~\cite{Hoang:2014wka}.
The updated profile functions have several advantages, and in particular exactly implement the canonical scale setting in the tail region. This improves the stability of the fit results to variations of the fit region, as can be seen in Fig.~\ref{fig:stability} when comparing the middle and left panels. Furthermore, the updated profiles have a variable slope and the parameters are in general more independent for different regions of the thrust spectrum.

\subsection{Impact of Resummation}\label{subsec:resummation}

It is interesting to analyse the impact of resummation, which pictorially can be seen when comparing the central and right panels of Fig.~\ref{fig:stability}. Varying the fit range by a small amount causes significant changes for fixed-order fits, which leads us to conclude that including resummation is very important to obtain stable fit results.

\section{Treatment of Power corrections and Fit Results}\label{sec:PowerCorrections}

In this section we first discuss the impact that including the model for 3-jet power corrections derived in Refs.~\cite{Caola:2021kzt,Caola:2022vea} has on the outcome of our fit.
We also introduce a physically motivated parametrization of the transition between regions where dijet and 3-jet power corrections dominate, which differs from Refs.~\cite{Caola:2021kzt,Caola:2022vea} (for further details see Ref.~\cite{Benitez:2024}).

\subsection{Effect of 3-jet renormalon model}\label{subsec:renormalonModel}

\begin{figure}
\begin{minipage}{0.33\linewidth}
\centerline{\includegraphics[width=1.1\linewidth]{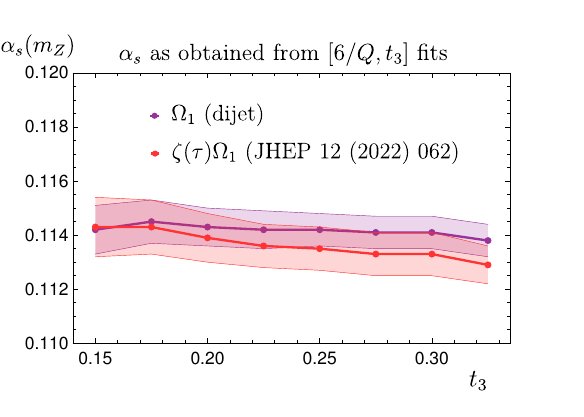}}
\end{minipage}
\hfill
\begin{minipage}{0.32\linewidth}
\centerline{\includegraphics[width=1.1\linewidth]{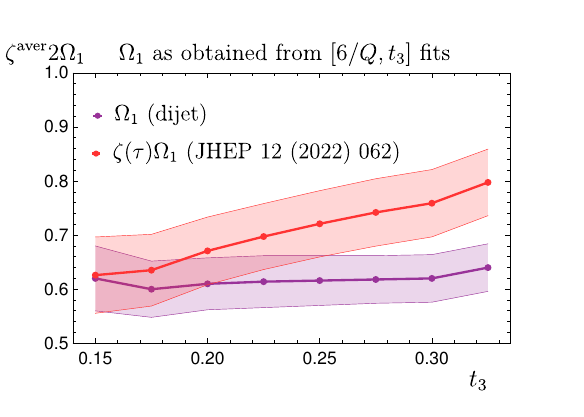}}
\end{minipage}
\hfill
\begin{minipage}{0.32\linewidth}
\centerline{\includegraphics[width=1.1\linewidth]{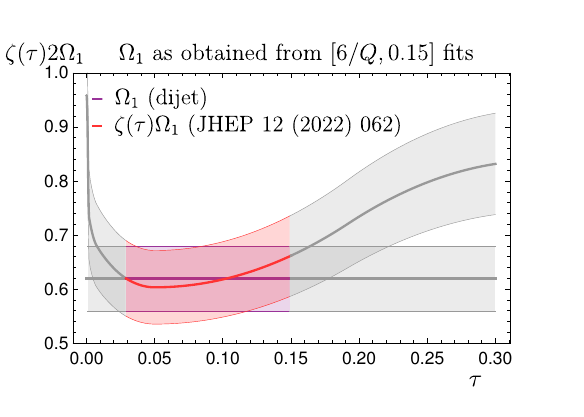}}
\end{minipage}
\caption[]{Comparison of the fit results obtained using a dijet power correction with those obtained when using the treatment of power corrections in the 3-jet region derived in Refs.~\cite{Caola:2021kzt,Caola:2022vea}. The left panel shows the result for $\alpha_s$ as a function of the upper bound of the fit range, $t_3$, whereas the middle panel illustrates $\zeta^{\rm aver} 2\Omega_1$ as a function of $t_3$. Here, $\zeta^{\rm aver}$ denotes the average values the model function from Refs.~\cite{Caola:2021kzt,Caola:2022vea} attains in the region where we perform the fit. The right panel depicts the function $\zeta(\tau)2\Omega_1$, where the corresponding best fit value for $\Omega_1$ comes from fitting in the range $\tau \in [(6\,{\rm GeV})/Q,0.15]$. The error bands contain only experimental uncertainties.}
\label{fig:powerCorr}
\end{figure}

In Fig.~\ref{fig:powerCorr} we compare fit results for $\alpha_s$ and $\Omega_1$ for two different treatments of the power corrections. On the one hand we use our previous treatment, which in the following we refer to as the dijet power correction (constant value for $\Omega_1$ across the entire spectrum). This is contrasted with the results we obtain using the renormalon model predictions of Refs.~\cite{Caola:2021kzt,Caola:2022vea}, giving $\zeta(\tau)2\Omega_1$ (our norm is such that $\zeta(0)=1$). From the left and middle panels of Fig.~\ref{fig:powerCorr} we observe that both treatments yield compatible results for the strong coupling, even if one goes far out in the distribution.
The results for $\Omega_1$ start to differ if we take the upper bound further out in the distribution, which is expected, since the 3-jet renormalon model develops a larger slope in this region, which will subsequently shifts the distribution. This effect is then compensated by a corresponding shift in $\zeta^{\rm aver}2\Omega_1$, but with much less impact on $\alpha_s(m_Z)$. The right panel most clearly illustrates the fact that, if we consider the upper bound of the fit range to be set to $t_3 = 0.15$, the two different treatments of power corrections largely coincide
when we compare $\zeta(\tau)2\Omega_1$ with the best fit value of $\Omega_1$
in this fit range.

\begin{figure}
\begin{minipage}{0.33\linewidth}
\vspace{0.25cm}\centerline{\includegraphics[width=0.925\linewidth]{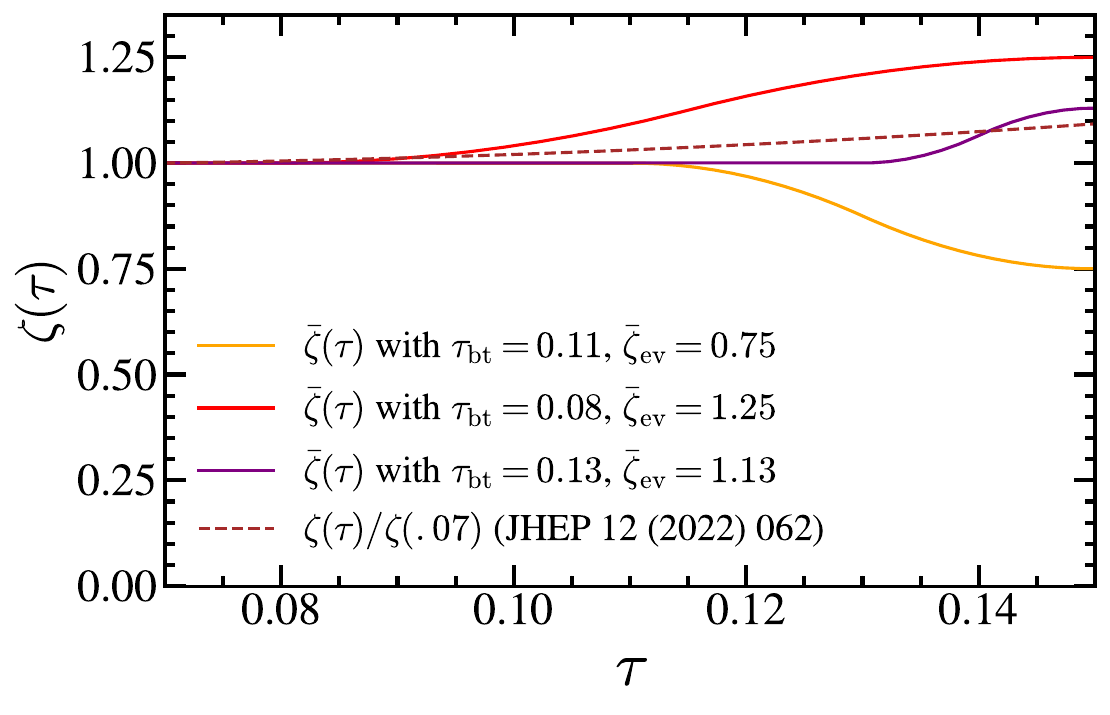}}
\end{minipage}
\hfill
\begin{minipage}{0.32\linewidth}
\centerline{\includegraphics[width=1.0\linewidth]{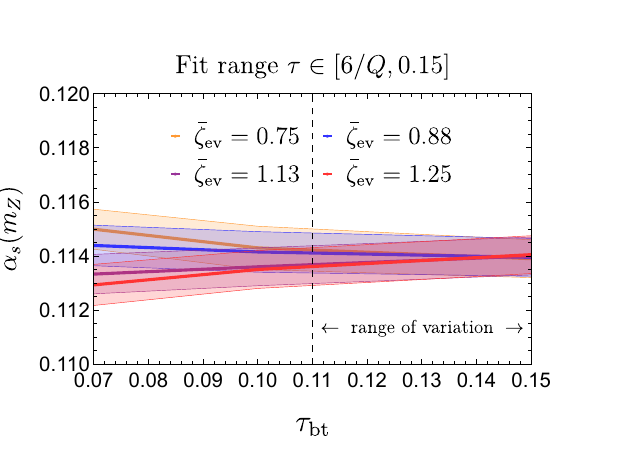}}
\end{minipage}
\hfill
\begin{minipage}{0.32\linewidth}
\centerline{\includegraphics[width=1.0\linewidth]{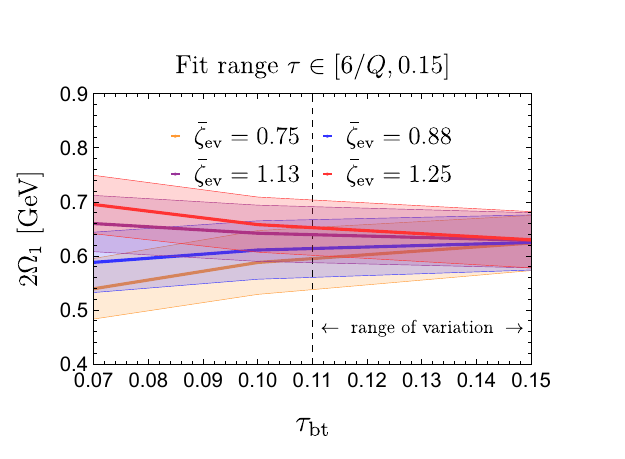}}
\end{minipage}
\caption[]{Left panel: Model function that parametrizes the effects of deviations from the dijet treatment for different parameter choices of $\tau_{\rm bt}$ and $\bar{\zeta}_{\rm ev}$. Center and right panels: Fit results when varying $\tau_{\rm bt}$ and $\bar \zeta_{\rm ev}$. The error bands contain only experimental uncertainties, while the fit range is fixed to $(6\,{\rm GeV})/Q \le \tau \le 0.15$.}
\label{fig:Ourmodel}
\end{figure}

\subsection{Parametrizing deviations from Dijet region}\label{subsec:deviationDijet}

Due to the newly obtained stability, depicted in the central panel of Fig.~\ref{fig:stability}, we will restrict our fit range to a region which
is clearly dijet dominated: $\tau \in [(6\,{\rm GeV})/Q,0.15]$.
Since this reduced fit window still might be affected by the transition from a dijet to a widely separated 3-jet final state, we assess the impact of such a transition as an additional source of uncertainty.
To model this transition, we consider the
parametrization $\bar\zeta(\tau)\Omega_1$ for the power correction, where the function $\bar\zeta(\tau)$ is shown in the left panel in Fig.~\ref{fig:Ourmodel} for a selection of different parameter choices. Our model function is based on the observation that dijet configurations with $\tau\ll 1$ clearly dominate up to a value of $\tau \sim 0.11$. Until this point in the spectrum, we consider the dijet prediction for the power correction, which corresponds to the flat line. Deviations from the dijet prediction are parametrized by a quadratic function that converges to a value $\bar{\zeta}_{\rm ev}$ for large $\tau$.
Varying the parameter responsible for the start of deviations away from the dijet treatment, $\tau_{\rm bt} $, as well as the value the model function attains at $\tau = 0.5$, $\bar{\zeta}_{\rm ev}$, provides us with an uncertainty band on the fit parameters shown in the middle and right panels of Fig.~\ref{fig:Ourmodel}. As our final uncertainty estimate, we take half of the difference between the maximum and minimum best fit
values of $\alpha_s$ (and $\Omega_1$, respectively), obtained when varying $\bar{\zeta}_{\rm ev}$ in the range $[0.75, 1.25]$, while fixing $\tau_{\rm bt} = 0.11$.

\subsection{Updated Fit results}\label{subsec:fitResults}

Performing a fit in the reduced fit range $[(6\,{\rm GeV})/Q,0.15]$, taking into account perturbative as well as experimental uncertainties, and adding in addition the error arising from the deviation of the dijet treatment of the power corrections, we obtain our preliminary result:
\begin{align}
\label{eq:finalFitResults}
\alpha_s(m_Z) &= 0.1142 \pm 0.0006_{\rm pert} \pm 0.0009_{{\rm exp}+\Omega_1} \pm 0.0004_{\rm had3j} = 0.1142 \pm 0.0012_{\rm tot}\,, \\[5pt]
\Omega_1 &= 0.313 \pm 0.033_{\rm pert}\pm 0.030_{{\rm exp}+\alpha_s} \pm 0.018_{\rm had3j}\,{\rm GeV} = 0.313 \pm 0.048_{\rm tot}\,{\rm GeV}\,, \nonumber\\[5pt]
\chi^2 / {\rm dof} &= 0.86\,.\nonumber
\end{align}
This value from our dijet focused analysis with a more conservative treatment of power corrections
is consistent with the earlier fit in Ref.~\cite{Abbate:2010xh} which used a wider fit range $[(6\,{\rm GeV})/Q,0.33]$.
Final results with a more complete discussion of uncertainties will be published in Ref.~\cite{Benitez:2024}.

\section{Conclusions}\label{sec:Conclusions}

We have revisited the determination of the strong coupling from fits to $e^+e^-$ event-shape data and found that resummation is important for the outcome of the fit, since it assures stability in terms of a variation of the fit window. Furthermore, different parametrizations of power corrections in the transition region to the 3-jet final state have been found to have a mild impact on the extraction of $\alpha_s$. At the light of our findings, it is safe to say that the small values found for the strong coupling in fits to the tail region of the thrust distribution with analytic power corrections cannot be attributed to the treatment of hadronization effects in the 3-jet region.

\section*{Acknowledgments}

VM, MBR and AH are supported in part by the Spanish MECD grant No.\ PID2022-141910NB-I00, the EU STRONG-2020 project under Program No.\ H2020-INFRAIA-2018-1, Grant Agreement No.\ 824093, and the COST Action No.\ CA16201 PARTICLEFACE. MBR is supported by a JCyL scholarship funded by the regional government of Castilla y Le\'on and European Social Fund, 2022 call.
I.S. is supported in part by the U.S. Department of Energy, Office of Nuclear Physics from DE-SC0011090 and the Simons Foundation through the Investigator grant 327942.

\end{document}